\documentclass{article}

\usepackage{bm}
\usepackage{amsmath}
\usepackage{amsfonts}
\usepackage{graphicx}
\usepackage{color}
\usepackage{amssymb}
\usepackage{hyperref}
\usepackage{caption}
\usepackage[ruled,linesnumbered]{algorithm2e}
\usepackage{multirow}
\usepackage{parskip}
\usepackage{tabularx}
\usepackage{graphicx,psfrag,fancyhdr,layout,appendix,subfigure}
\usepackage{verbatim}
\usepackage{framed}
\usepackage{amsthm}
\usepackage{booktabs}
\usepackage{placeins}
\usepackage{rotating}
\usepackage{changebar}
\usepackage{dcolumn}
\usepackage{multirow}
\usepackage{times}

\makeatother			

\newcommand{\manycites}[1]{\cite{#1}}
\newcommand{\fewcites}[1]{}

\begin{document}

\title{Local and Distributed Quantum Computation}

\author{Rodney Van Meter and Simon J. Devitt}

\maketitle

\begin{abstract}
  Experimental groups are now fabricating quantum processors powerful
  enough to execute small instances of quantum algorithms and
  definitively demonstrate quantum error correction that extends the
  lifetime of quantum data, adding urgency to architectural
  investigations.  Although other options continue to be explored,
  effort is coalescing around topological coding models as the most
  practical implementation option for error correction on realizable
  microarchitectures.  Scalability concerns have also motivated
  architects to propose distributed memory multicomputer
  architectures, with experimental efforts demonstrating some of the
  basic building blocks to make such designs possible.  We compile the
  latest results from a variety of different systems aiming at the
  construction of a scalable quantum computer.
\end{abstract}

\section{Introduction}

Quantum computers and networks look like increasingly inevitable
extensions to our already astonishing classical computing and
communication
capabilities~\fewcites{ladd10:_quantum_computers,van-meter14:_quantum_networking}~\manycites{nielsen-chuang:qci,van-meter14:_quantum_networking}.
How do they work, and once built, what capabilities will they bring?

Quantum computation and communication can be understood through seven
key concepts (see sidebar).  Each concept is simple, but collectively
they imply that our classical notion of computation is incomplete, and
that quantum effects can be used to efficiently solve some previously
intractable problems.

In the 1980s and 90s, a handful of algorithms were developed and the
foundations of quantum computational complexity were laid, but the
full range of capabilities and the process of creating new algorithms
were poorly
understood~\manycites{nielsen-chuang:qci,deutsch85:_quant_church_turing,feynman:_simul_physic_comput,bennett:strengths,bernstein97:_quant_complexity}.
Over the last fifteen years, a deeper understanding of this process
has developed, and the number of proposed algorithms has
exploded~\footnote{{\tt http://math.nist.gov/quantum/zoo/}}.  The
algorithms are of increasing breadth, ranging from quantum chemistry
to astrophysics to machine learning-relevant matrix
operations~\fewcites{montanaro2015:qualgo-qi}\manycites{mosca2008quantum,bacon10:_recent_progress,montanaro2015:qualgo-qi}.
Some algorithms offer only a polynomial speedup over competing
classical algorithms; others offer super-polynomial speedups in
asymptotic complexity.  However, many of these algorithms have not yet
been analyzed in an architecture-aware fashion to determine constant
factors, fidelity demands and resource requirements.  The required
size, speed and fidelity of a commercially attractive computer
remains an open question.

To the DiVincenzo criteria (see box) we have added a number of
practical engineering constraints: for example, systems must be small
enough, cheap enough, and reliable enough to be practical, and fast
enough to be useful.  Due to implementation limitations, it is
imperative for scalability that locally distributed computation be
achievable, which requires system area networks that are fast, high
fidelity, and scalable.

Beyond tight coupling of small quantum computers into larger
multicomputers to scale monolithic algorithms lies the realm of
distributed quantum algorithms and sensing.  These applications, which
bridge pure numerical computation with cybernetic uses, will improve
the sensitivity and accuracy of scientific instruments, as well as
augment classical cryptographic capabilities.

\hrule
\section*{\emph{sidebar:} Quantum Computing Concepts}

Many quantum phenomena exhibit a set of discrete states,
such as the energy levels of an atom, the direction of spin of an
electron (aligned or anti-aligned to the local magnetic field), or
horizontal and vertical light polarization.  We begin by selecting two
separate, orthogonal states of one of these phenomena to be the zero
and one states of our \emph{qubit}.

\emph{Superposition} in quantum systems acts in a somewhat analogous way to
classical wave mechanics.  Light polarized at a 45 degree angle is an
even superposition of horizontal and vertical polarization.  Less
obviously, we can also create superpositions of any of our qubit
candidates, including two electron spin states or two atomic energy
levels.  The probability that, in the end, a certain outcome will be
found is related to the relative amounts of zero and one we put into
the superposition.

When we have more than one particle or qubit in our quantum system, we
cannot in general talk about their states independently, because the
qubits can be \emph{entangled} in such a way that the state of each
depends on the other.  This correlation is stronger than dependent
classical probabilities, and forms the basis of quantum communication.
It is important to note that entanglement cannot be used to
communicate faster than the speed of light, even though entangled
particles that are far apart will show correlations with no classical
explanation when used appropriately.

As our system grows, $n$ qubits have $2^n$ possible states, 0...0 to
1...1, just as with classical bits; we call the set of qubits our
register.  Our total state is described by the wave \emph{amplitude}
and \emph{phase} of each of these states, thus a complete classical
description of a state can require as much as $O(2^n)$ memory.

The quantum algorithm designer's job is to shuffle amplitude from
value to value, altering the superposition while manipulating the
phase to create \emph{interference}: constructive interference, when
phases are the same, increases the probability of a particular
outcome, while destructive interference, when phases differ, reduces
the probability.

In a circuit-based quantum computer, the algorithm designer's tool is
\emph{unitary}, or \emph{reversible}, evolution of the state.  She
does this by defining a series of gates that change one or two qubits
at a time, roughly analogous to classical instructions or Boolean
logic gates.  The controlled-\textsc{not}, or \textsc{cnot}, is one
such common building block.

A significant exception to reversibility is \emph{measurement}, in
which we look at the quantum system and extract a numeric value from
the register.  This causes the collapse of the superposition into a
single state.  Which state the system chooses is random, with
probabilities depending on the relative amplitudes of different
states, taking interference into account.  Measurement destroys
entanglement.

Quantum states are very fragile, and we must keep them well isolated
from the environment.  However, over time, errors inevitably creep in,
a process known as \emph{decoherence}.  The natural classical solution
would be to keep extra copies of fragile data, but the
\emph{no-cloning theorem}, a fundamental tenet of quantum mechanics,
dictates that it is not possible to make an independent copy of an
unknown quantum state.

\hrule
\hrule
\section*{box: DiVincenzo Criteria}

As researchers began to feel their way through the notion of a quantum
computing machine, David DiVincenzo laid out a set of five criteria a
quantum computing technology would have to meet in order to build a
computer, later clarified with the addition of two more, for
communication~\manycites{divincenzo2000piq}.

\begin{itemize}
\item First, we must have an \emph{extensible register of two-level
    systems}, usable as qubits.  The simple word \emph{extensible}
  hides substantial engineering complexity, addressed in the main
  text.
\item Second, the register must be \emph{initializable to a known
    state}.
\item Third, we must have \emph{universal gate set}, the ability to
  achieve any proposed algorithm that fits within the basic framework
  of quantum computation.
\item Fourth, our qubits and operations on them must exhibit
  \emph{adequate coherence time and fidelity} for long quantum
  computations.  Early criticism of quantum computation centered
  around this fact~\manycites{unruh95:_maintaining-coherence}, leading to
  the development of quantum error correction and fault tolerance
  (main text).
\item Fifth, a computer from which we can't extract the results is not
  very useful, so we must have \emph{single-shot measurement}.
\end{itemize}

With the above, we can construct a basic quantum computer.  In order
to achieve scalability through photonic interconnects, or to create
networks delivering entangled states to applications, we also need the
ability to \emph{convert between stationary and flying qubits
  (photons)}, and we need to be able to \emph{capture and control the
  routing of our photons}.

\hrule

\section{Architectural Models for Large-Scale Computation}
\label{sec:architectures}

Theoretical architectures for large scale quantum computation now
almost exclusively rely on topological models of error correction
\manycites{K03}, with the surface code \manycites{FMMC12} and the
Raussendorf code \manycites{RHG06} now dominating
designs~\fewcites{fowler:PhysRevA.86.032324}.  Each of these systems
utilizes a different physical technology that defines the qubit and
adapting error correction models to the physical restrictions of
quantum hardware has led to multiple architecture designs, indicating
a clear pathway towards future quantum computers
\manycites{DFSG08,JMFMKLY10,YJG10,NTDS13,MRRBMD14,AMK15,LWFMDWH15,LHMB15,HPHHFRSH15,ONRMB16}.

The surface code and Raussendorf code have been adopted broadly for
three reasons.  Memory lifetimes and gate error rates remain a
challenge for experimentalists, and the surface codes have high
thresholds, approaching 1\% depending on the
physical model \manycites{S14}.  The intrinsic nearest neighbor
structure ensures that the physical hardware does not require long
range interactions~\manycites{CWCW13}.  The software driven programming
model for manipulating logical qubits allows run-time allocation of
resources to suit any application algorithm, including (within limits)
adjustment of the error correction strength \manycites{D14,PPND15,PDF16}.
One perceived drawback is the high resource cost, with many physical
qubits per logical qubit, but the analyses suggesting large numbers
were conducted assuming physical error rates above the operational
thresholds of other codes~\manycites{van-meter10:dist_arch_ijqi,van-meter13:_blueprint}.  

Topological coding models allow architectures to be designed with a
high level of modularity.  Small repeating elements plug together to
form a computer of arbitrary scale; we refer to the architecture of
the unit for executing error correction as the microarchitecture.  The
comparative simplicity of the hardware structure makes it far easier
to experimentally build and currently the biggest challenge is
engineering qubit components with the required fidelity for
topological error correction to become effective.

\begin{figure*}
\begin{center}
\resizebox{\linewidth}{!}{\includegraphics{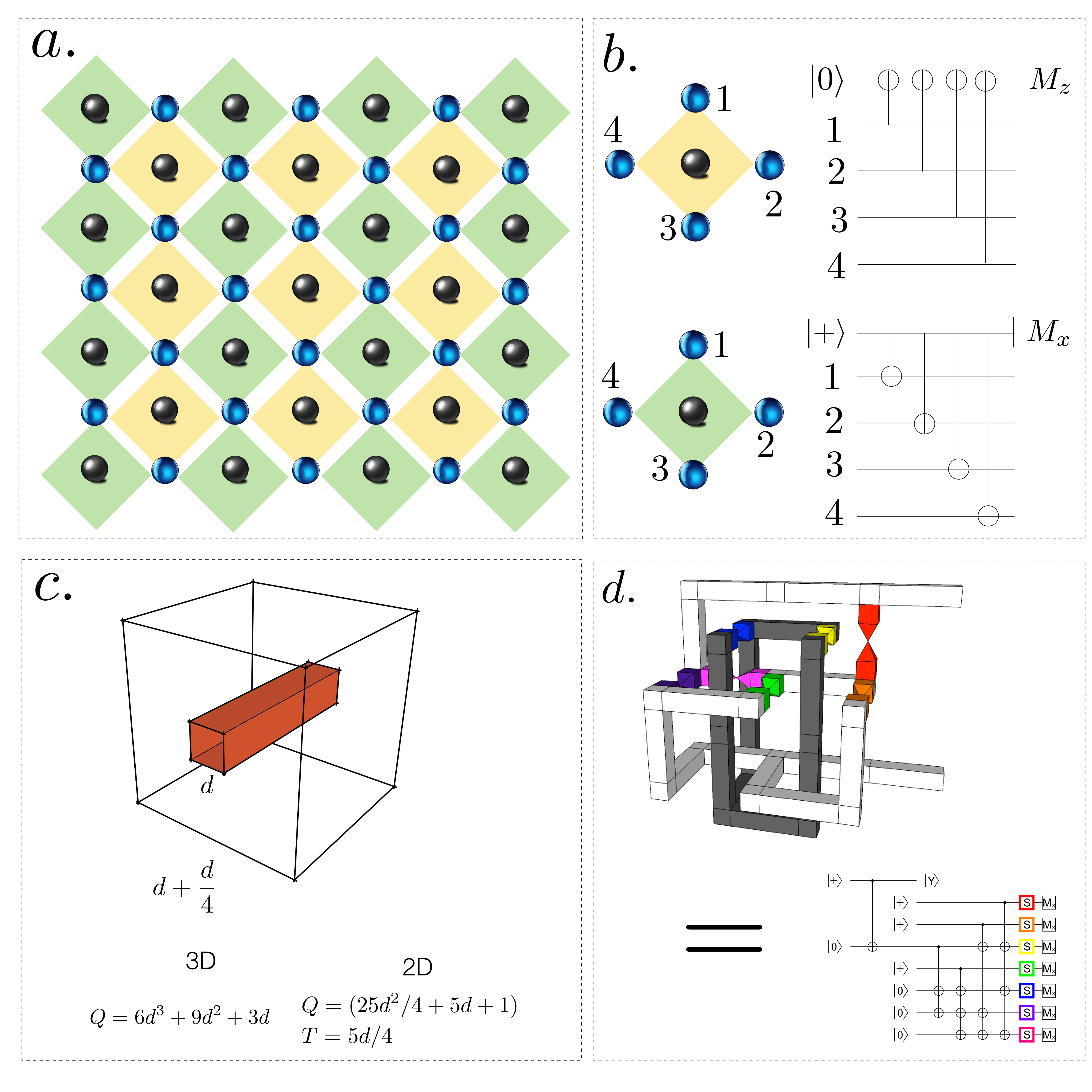}}
\end{center}
\caption{{\bf Elements of the surface code.} Figure a. 
  is the 2D structure, with blue dots representing data qubits and
   black dots representing syndrome qubits for error correction
  information.  Figure b. illustrates the two types of circuits that
  are continuously run across the lattice to detect errors.  Figure c.
  illustrate a {\em plumbing piece} which is a fundamental element of
  a topological quantum circuit \manycites{FD12}.  
  The size of the plumbing piece is related to a scale factor of
  lattice spacings, $d$, where $d$ is the error correction strength of
  the topological code which will tell you the number of qubits
  required for an implementation in the Raussendorf lattice
  \manycites{RHG06} or the number of qubits and time steps for the
  surface code \manycites{FMMC12}.  Figure d. illustrates an optimized
  quantum circuit in this model \manycites{FD12}.}
\label{fig:surface}
\end{figure*}

Several detailed reviews of the topological coding model cover the
functioning of both the error correction and logical
computation~\manycites{FMMC12,LB13,F15,horsman2012surface-njp}.  While the model is complicated,
the basic hardware configuration is quite simple.  
The 2D surface code is illustrated in Figure \ref{fig:surface}a.  Half
of the qubits are {\em data} qubits that are part of the code (blue) and the
other half of the qubits are {\em syndrome} qubits that are used to
extract error information and act as an entropy sink for the encoded
data (black).  The two circuits shown in Figure \ref{fig:surface}b 
are run continuously across the entire computer in parallel and subsequent
measurements of each syndrome qubits are used to identify physical
errors across the surface.

In the surface code model, computation is achieved by temporarily {\em
switching off} some of these circuits, creating holes ({\em defects})
within the surface.  These defects introduce a degree of freedom
within the system which we use as a logical qubit, protected against
errors by being physically large in cross-section and separated from
other defects (or boundaries) in the lattice.  As the size and
separation of these defects increase linearly, the logical error rate
of the information drops off exponentially.  This switching on or off
regions of the computer to create and manipulate encoded defects is
what allows us to translate a compiled fault-tolerant quantum circuit
(see Section \ref{sec:software}) into the physical control signals of
the computer (Figure \ref{fig:surface}b,c).  

The Rausesendorf model of topological computation \manycites{RHG06} uses a
large, entangled state known as a cluster state, arranged in a 3D
lattice.  It continually entangles new qubits into one surface of the
state and measures older qubits from the opposite face, removing them
from the state, in a rolling fashion.  Each time slice in the 2D
surface code is now represented along the third dimension.
Information is continuously teleported along this third dimension with
both error correction and data processing occurring via these
teleportations.  While the Raussendorf and surface code models are
formally equivalent (and programming both models is essentially the
same), hardware concerns make one model sometimes more preferable.  A
general rule of thumb is that the surface code is more appropriate for
hardware where individual qubits are physically static and are able to
be measured without destroying the physical qubit
\manycites{JMFMKLY10,LWFMDWH15,HPHHFRSH15}, while the 3D Raussendorf code
is reserved for ``flying qubit'' technologies (most notably optics)
and for some architectures where probabilistic gates are heavily used
\manycites{NTDS13,LHMB15}.

Designs for large-scale quantum computers predating the development of
the surface codes relied on multiple layers (concatenation) of
classically derived quantum error correcting codes, and some
researchers continue to pursue this
approach~\manycites{ahsan2015designing}.  These older codes are simpler to
decode at run time, and if the underlying technology supports
long-distance interaction between qubits, the primary technical
challenge is the higher required fidelity for physical operations.

Because the scale of complete systems will be large, and the physical
sizes of qubit structures are far larger than transistors, it has
become common to assume the macroarchitecture of the system will be a
multicomputer design.
\section{Experimental Progress}

Since the landmark 2010 review from Ladd {\em
  et. al.}~\fewcites{ladd10:_quantum_computers}
\manycites{ladd10:_quantum_computers}, experimental progress towards
large-scale quantum computers warrants an update, in the context of
scaling systems.  What will constrain our ability to build a quantum
computing system as large as we care to attempt?  Van Meter's thesis
offered the following broad but informal definition of
scalability~\manycites{van-meter06:thesis}:

\begin{quotation}
Above all, it must be possible, physically and economically, to grow
the {\em system} through the region of interest.  Addition of physical
resources must raise the performance of the system by a useful amount
(for all important metrics of performance, such as calculation speed
or storage capacity), without excessive increases in negative features
(e.g., failure probability).
\end{quotation}

This definition refers to several important criteria.  It also points
out that scalability is never indefinite in the real world.  No one
would say that a system that costs a hundred thousand dollars per
qubit or with m$^2$ gate footprints is scalable in
any practical sense.  Systems that can be built, but only for
exorbitant costs (e.g., above a billion dollars), or require
unavailable quantities of helium or other resources may be scalable on
paper but not in the real world.

\begin{figure*}
\begin{center}
\resizebox{\linewidth}{!}{\includegraphics{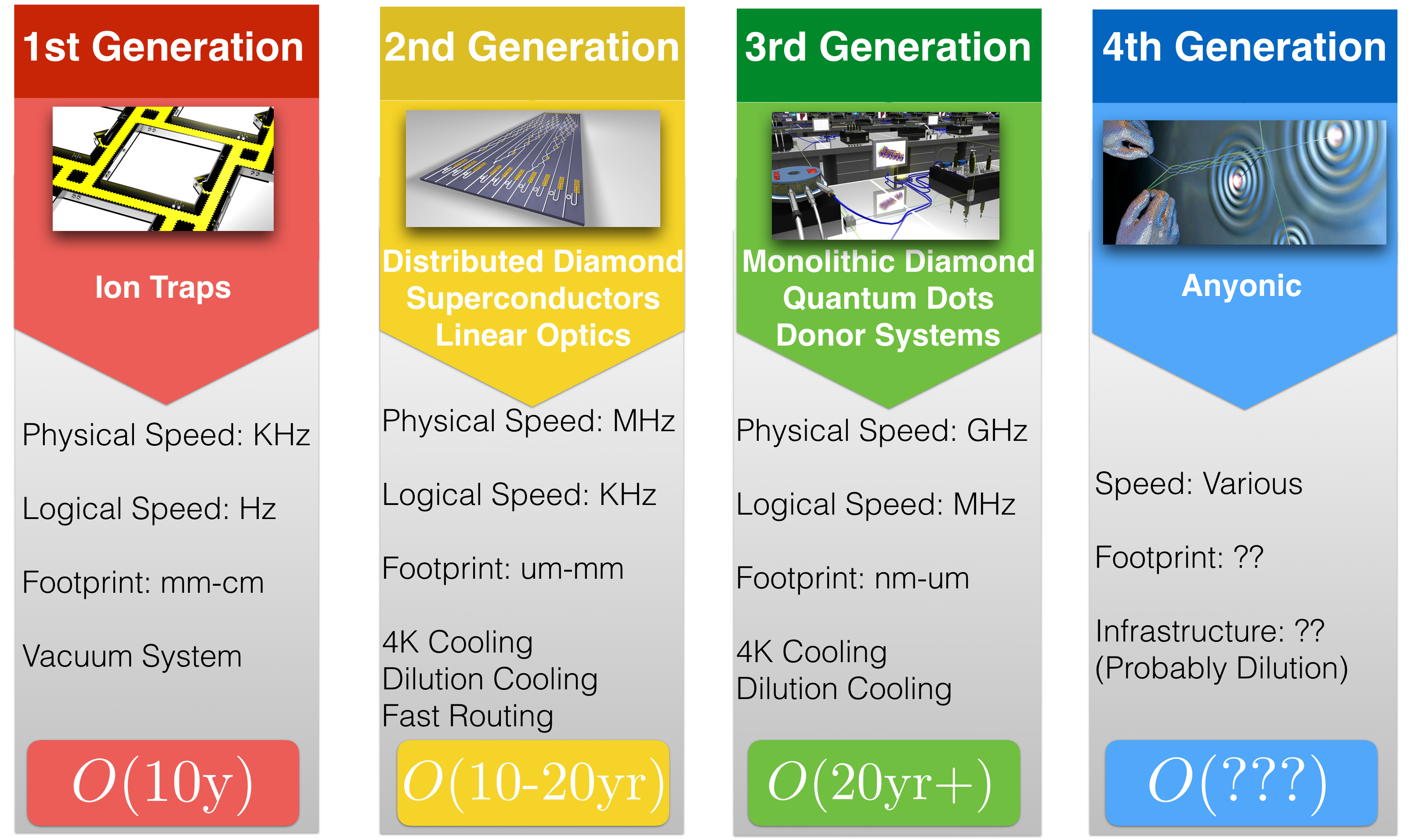}}
\end{center}
\caption{{\bf Generations of Quantum Computers.} Qualitative
  assessment of the seven major quantum computing technologies (specific implementations 
  may straddle generations in one or more metrics).  Each
  physical system has its place within a broader industrial sector in
  quantum computing.  In general, systems that have the potential to
  be smaller, faster and cheaper are less developed than systems that
  will be big, slow and costly.  
  }
\label{fig:generations}
\end{figure*}

We are of the opinion that differing quantum computing technologies
are complementary and have a well defined place within an emerging
technology sector.  Developmental timeframe, cost, execution speed and
physical size are metrics that can vary by orders of magnitude between
systems and generally the systems that potentially offer higher
performance are less developed.  A qualitative summary of seven major
technologies receiving significant academic and
industrial attention is illustrated in Figure \ref{fig:generations}.

\subsection{Ion Traps}

Ion trap quantum computing was an early experimental success
story~\fewcites{brown2016co}\manycites{brown2016co}.  This success can
be largely attributed to technological developments in ion trapping
motivated by uses such as atomic clocks.  An ultrahigh vacuum is used,
and individual atoms are ionized and ``trapped'', held in place and
controlled using carefully controlled electrical fields.  Cirac and
Zoller first proposed quantum computing using trapped ions in
1995~\manycites{CZ95} and the demonstration of primitive gate
operations occurred soon afterwards \manycites{MMKIW95,LBMW03}.  A
large-scale quantum computer with all of the qubits in a single trap
is impractical for several reasons, such as slower gate times, cross
talk when applying quantum gates, limited operational parallelism, and
increases in decoherence rates.

To combat these problems, the idea of segmented traps was proposed
\manycites{KMW02}.  This microarchitectural model uses a series of DC
electrodes that can move the electrostatic potential along a trapping
pathway, essentially dragging the ion with it. 
Individual qubits can be placed into storage regions and then moved to
interaction zones for multi-qubit gates.
This segmented design requires delicate control to ensure that ions
can be moved without losing them around complex trapping geometries
\manycites{H06,BOVABLW09}.

A simple approach to a large system is a monolithic design
\manycites{LWFMDWH15} where individual segmented ion traps are
fabricated, aligned and connected together to form the complete
computer.  This design has the advantage that physical operations for
surface code are as simple as possible.  The downside to this design
is the need for vacuum infrastructure surrounding the entire computer
and the sheer size of a machine containing millions or billions of
qubits (footprint estimates are approximately
6mm$^2$/qubit~\manycites{LWFMDWH15}).

A second approach is to further divide the computer into small
Elementary Logic Units (ELUs) \manycites{MRRBMD14,AMK15}.  Each ELU may be
a segmented trap holding tens to thousands of qubits.  They are
interconnected using probabilistic optical connections achieved by
optically exciting two distant ions and using the emitted photons
\manycites{DMMMKM06}.  This communications channel allows the connection of
independent ELUs to form a larger multicomputer.

While this approach mitigates the infrastructure issues that would
plague a monolithic ion trap computer, it does introduce
complications.  The optical connections that allow the creation of
entanglement between ions are intrinsically probabilistic.  Coupled
with inefficiencies in capturing emitted photons, detector
inefficiencies and loss through optical switches, many attempts are required before a
successful connection is achieved.  Initial experiments required on
the order of tens of minutes to establish entanglement between ions
\manycites{DMMMKM06}, but this has improved to on the order of five times
per second~\manycites{hucul2014modular}.  Architectural structures have
been proposed for these optically connected ion-traps based on both
topological codes and traditional concatenated codes
\manycites{MRRBMD14,AMK15}, however the rate of the optical connections
needs to be increased.

Ion trap systems are progressing rapidly and may represent the first
physical system to outperform classical quantum computers.  However,
the size, speed and potential cost of an ion trap quantum computer may
restrict the ultimate scalability.

\subsection{Superconductors}

Superconducting quantum computers have seen explosive success in the
past five years and are the principal technology of two of the first
industrial players in the quantum
sector~\fewcites{martinis2015qubit-npjqi}.  Both the group of John
Martinis, now at Google inc. and the IBM effort are utilizing
superconducting qubits and surface code techniques to push forward in
building large-scale systems.  Superconducting qubits come in several
flavors; the most successful variants use a quantized amount of
current in a loop of superconductor.  They can be considered a
generation beyond ion traps as they have intrinsic gate times on the
order of 100ns \manycites{B14} (CZ gates are about twice as fast) and
have qubit footprints on the order of approximately 100 microns
square~\manycites{D09}.  Superconducting qubits have demonstrated
single qubit, two-qubit, initialization and measurement error rates
below the fault-tolerant threshold within a single
device~\manycites{B14}.

Several proposed architectures now exist for large-scale quantum
computers~\manycites{FMMC12,B16}.  Monolithic approaches from Google and
IBM have illustrated the necessary building blocks for fault-tolerant
computation using the surface code \manycites{B14,C15}, but a major
challenge is to scale to a 2D nearest neighbor array of qubits while
not degrading the individual error rates of qubits.  IBM demonstrated
a 2x2 array of superconducting qubits \manycites{C15}, but larger arrays
will be needed and the fabrication and placement of the necessary
control wiring for each qubit is an engineering problem that has yet
to be solved.  As with ion traps, distributed designs have the
potential to mitigate infrastructure and control issues for a
large-scale machine but introduce more complicated protocols in order
to realise the fundamental gate libraries necessary to implement
either surface code or other error correction techniques across a
slower, more error-prone interconnect~\manycites{B16}.

The rapid advances in the past few years now raise the very strong
possibility that superconducting quantum computers, rather than ion
traps, may be the first to realize a digital quantum computer that can
outperform classical machines on relevant quantum problems.  Besides
IBM and Google, startups such as Rigetti computing
\footnote{www.rigetti.com} and Quantumcircuits,
Inc. \footnote{www.quantumcircuits.com} are now specifically targeting
this platform.

\subsection{Linear Optics}

Linear optics was one of the first platforms to demonstrate the
building blocks of quantum
computation~\fewcites{Bonneau2016}\manycites{Bonneau2016}.  The
initial theoretical foundation for linear optical quantum computing
can be arguably attributed to the seminal paper of Knill, Laflamme and
Milburn \manycites{KLM01}, who showed that measurement induced
non-linearities and hence universal computation was possible.

After initial demonstrations of the building blocks of linear optical
quantum computation in bulk optics \manycites{OPWRB03,POWBR04},
development moved into the field of integrated optics, where
individual photons are routed through etched waveguides in a bulk
material (ostensibly silicon) \manycites{P08}.  Early efforts were
extremely successful, with high fidelity circuits performing small
quantum programs \manycites{O07,O09}.  More recent effort has focused
on integrating all aspects of a universal quantum computing system
(the photon sources, detectors and waveguides) on chip
\manycites{SB14,S15,SS15}.  High fidelity, high efficiency, on-demand
single photon sources still remain the Achilles heel of the
technology.  There are generally two approaches: using an atomic based
photon emitter to produce on demand-photons
\manycites{KH13,SG16,CB10,GM13}, or using Spontaneous Parametric Down
Conversion sources (SPDCs) and optical switching to create a
multiplexed source \manycites{MZ11,MSM15}.  Using multiplexed sources
translates the source problem into constructing very low loss,
single-photon switches \manycites{BMO15}, which is the focus of
current research.

Architecture for linear optics has also
progressed with two general approaches for realizing a topologically 
protected machine.  The first is to slowly construct a 3D
Raussendorf lattice by probabilistically fusing together larger and
larger components \manycites{LBS10,LHMB15}.  This approach has the
downside of requiring the optical storage of the cluster
as it is grown and routing together smaller sub-clusters that may
have been successfully prepared in distant physical locations in the computer.  
This type of architecture has high overhead and non-trivial routing issues
requiring very low loss single photon switches.  The second approach 
is known as the ballistic model, where photons are sent through a fixed network 
of fusion gates and an incomplete ({\em Swiss cheese}) graph state is produced
\manycites{MSBR15}.  This model relies on the probability of success for
individual fusion gates being above approximately 63\% \manycites{MSBR15} 
(utilizing a technique known as Boosted Fusion \manycites{G11}) such that the 
created lattice percolates.  This approach relaxes the routing and storage 
requirements, but replaces it with taxing classical 
computational costs to calculate how to convert the {\em swiss cheese} 
lattice into a perfect Raussendorf lattice, in real time.  This second model has 
still not been fully analyzed and so overall resource overheads are unclear.

Linear optical quantum computers still have significant potential, but both
theoretical and experimental work is incomplete.  The
benefits of comparatively low infrastructure costs are a
significant selling point for the technology.

\subsection{Diamond}

Impurities in diamond have long been of interest as a potential
technology for both large-scale quantum computing and
communications~\fewcites{greentree16:diamond}\manycites{greentree16:diamond}.
The Nitrogen Vacancy (NV) center is by far the most commonly
researched type for potential use in active quantum
technologies~\manycites{PG08,LJLNMO10}.  Diamond is of interest due to
the ability to couple the NV center with a photon at optical
frequencies.  This allows for a natural interconnection between
stationary qubits (used as quantum memories) and flying qubits (used
on communications links).

Diamond based quantum computing architectures have been proposed both
in monolithic crystals, where numerous NV centers are fabricated
within a single crystal \manycites{YJG10} and more distributed
optically connected diamond arrays \manycites{NTDS13}.  As with
essentially all modern architectures, both of these proposals are
based upon the surface code or Raussendorf model.  While diamond has
experimentally demonstrated various elements required for large-scale
computation \manycites{TC10,NB10,RC11,MK12,BH13,DJ13}, researchers
have not yet shown high enough fidelity operations or a universal set
of gates within a single device.

Diamond based qubits were used to violate the famous Bell inequalities
in the first loophole free test in 2015 \manycites{HBD15}.  An ensemble
array of NV-centers was successfully coupled to a superconducting flux
qubit in 2011 \manycites{ZS11}.  In this system, the diamond layer is envisaged as a
method to couple superconducting qubits (which themselves would couple
via microwave photons) via optical photons.

While diamond based quantum computers are not as well developed as
other systems, several research groups are focused on this technology.
Diamond does not require as stringent infrastructure costs as ion
traps or superconductors.  Vacuums are not needed and cooling can be
limited to 4K, rather than millikelvin temperatures.  Lower potential
infrastructure costs and fast operation times makes diamond an ideal
bridge between 2nd and 3rd generation quantum technologies.

\subsection{Quantum Dots}

Quantum dots trap individual electrons at the boundary between
different semiconductor materials, and can be controlled optically,
electrically or magnetically~\fewcites{zwanenburg:RevModPhys.85.961}\manycites{zwanenburg:RevModPhys.85.961}. They are arguably not as experimentally
advanced as superconductors or ion traps, in large part due to
sensitivity to noise, which is now being
overcome~\manycites{reed:PhysRevLett.116.110402}.  As with donor systems,
they have the potential for denser integration and fast operation, but
this conceptual advantage is tempered by the apparent need for more
development time.  Quantum dots have many uses besides quantum
computing, including sensing, communications, and classical computing,
but a number of groups worldwide are working toward quantum dot
quantum computers.

Motivated by the original 1998 architecture of Loss and DiVencenzo
\manycites{LD98}, progress has been substantial.  Some of the quantum dot
groups do not fall within the academic sector, and limit their public
information, hence their progress is difficult to assess.  As with
each technology we have discussed, researchers have assumed a
large-scale structure based on topological surface codes
\manycites{JMFMKLY10,van-meter10:dist_arch_ijqi}.  

Experimental demonstrations of building block protocols have also been
pronounced.  Addressable quantum dot qubits with fault-tolerant levels
of control fidelity have been demonstrated \manycites{VD14,TT16} along
with a full two-qubit logic gate \manycites{VD15}.  Like the other six
major quantum computing systems, quantum dots are one of the more
promising systems for producing fast, small and cheap qubits.
However, further experimental development is needed to demonstrate
required building blocks of a scalable machine.

\subsection{Donors}

Donor based quantum computing systems use semiconductor dopants that
provide an extra, unpaired
electron~\fewcites{Hill:e1500707}\manycites{Hill:e1500707}.  In room
temperature semiconductor operation, the extra atom moves through the
material, but in quantum computing systems, the material is cooled to
millikelvin temperatures and the electron remains bound to its dopant
atom.  The goal is to use these individual electrons as spin qubits,
sometimes in conjunction with the nuclear spin of nearby atoms.  These
systems are exemplified by the P:Si technology, which has shown
significant experimental progress in the past five years
\manycites{ZDM13}.  The original 1998 architectural proposal made by
Kane did not consider the challenges of error correction or
algorithmic implementation \manycites{K98}.  Since then, several
generations of architectures for P:Si quantum computers have been
proposed~\manycites{HGFW06,HPHHFRSH15,ONRMB16}.

Experimentally, there were significant challenges to simply build a
functional qubit using phosphorus donors, as an atomically precise
array of phosphorus donors needs to be embedded within an otherwise
isotopically pure crystal of $^{28}$Si.
The actual placement of
the phosphorus donors within the crystal followed two methods, known
as Top Down \manycites{DHJ03} and Bottom Up \manycites{SCS03}.  Top Down
involved direct injection of the phosphorus via a focused ion beam.
Direct injection is not atomically precise and causes significant damage to the silicon
substrate that needs to be annealed (which can also cause donors to
move).  The Bottom Up approach grows, layer-by-layer, the
Silicon substrate and then places, with atomic precision, each
phosphorus donor and then continues to grow the silicon layer on top.
This method is more precise and is now preferred for scalable fabrication.

Since 2010, P:Si technology has progressed from readout and
addressability of small clusters of phosphorus donors \manycites{BMR13}, to
demonstrating the anticipated long coherence time of a single donor in
isotopically pure silicon \manycites{MDJ14}, high fidelity readout
\manycites{PTK13}, single qubit control \manycites{PTDJ12} and violations of a
Bell inequality using the electron and nuclear spin of a single
phosphorus donor \manycites{SSM16}.

The original motivation of leveraging the technology in the classical
silicon industry remains strong.  While it is expected that other
technologies will achieve a large-scale machine earlier, donor based
quantum computers are an attractive option due to the potential for
donor systems to be smaller and cheaper.

\subsection{Anyons}

Anyonic quantum computing is often referred to as topological quantum
computing, for good
reason~\fewcites{das-sarma2015:majorana-npjqi}\manycites{das-sarma2015:majorana-npjqi}.
The original description of the toric code \manycites{K03} very
quickly recast an otherwise quantum error correction coding mechanism
into a Hamiltonian formulation and postulated the existence of quantum
particles that exhibit fractional quantum statistics (in contrast to
the usual integer statistics of Bosons and Fermions).  We use the
terminology {\em anyonic quantum computers} to distinguish this model
from the topological coding models that have already been discussed.

We cannot provide a complete review of the field here, as it has
emerged as a very complicated model of quantum computation.  There are
already excellent summaries of both the theoretical foundations
\manycites{FKLW03,NSS08} and possible implementations \manycites{SDN15}.  As
illustrated in Figure \ref{fig:generations}, we have assigned anyonic
quantum computing to the 4th generation, for two reasons.  First,
anyonic quantum computing tries to suppress errors using the
fundamental physics of the system itself.  Rather than embed
complicated error correction codes on top of standard two-level
quantum systems (qubits), the idea is to engineer a system that
exhibits quantum excitations that are naturally protected from
decoherence.  This may lead to systems with extremely low physical
error rates, mitigating (or even eliminating) the need for active
error correction.  The second reason for casting anyonic computing as
a 4th generation system is that we need to reliably demonstrate the
existence of anyonic particles within engineered
systems~\manycites{MZFK12,AHM16}.

\section{Progress in Software Control for Large-Scale Computation}
\label{sec:software}

\begin{figure*}[ht!]
\begin{center}
\resizebox{\linewidth}{!}{\includegraphics{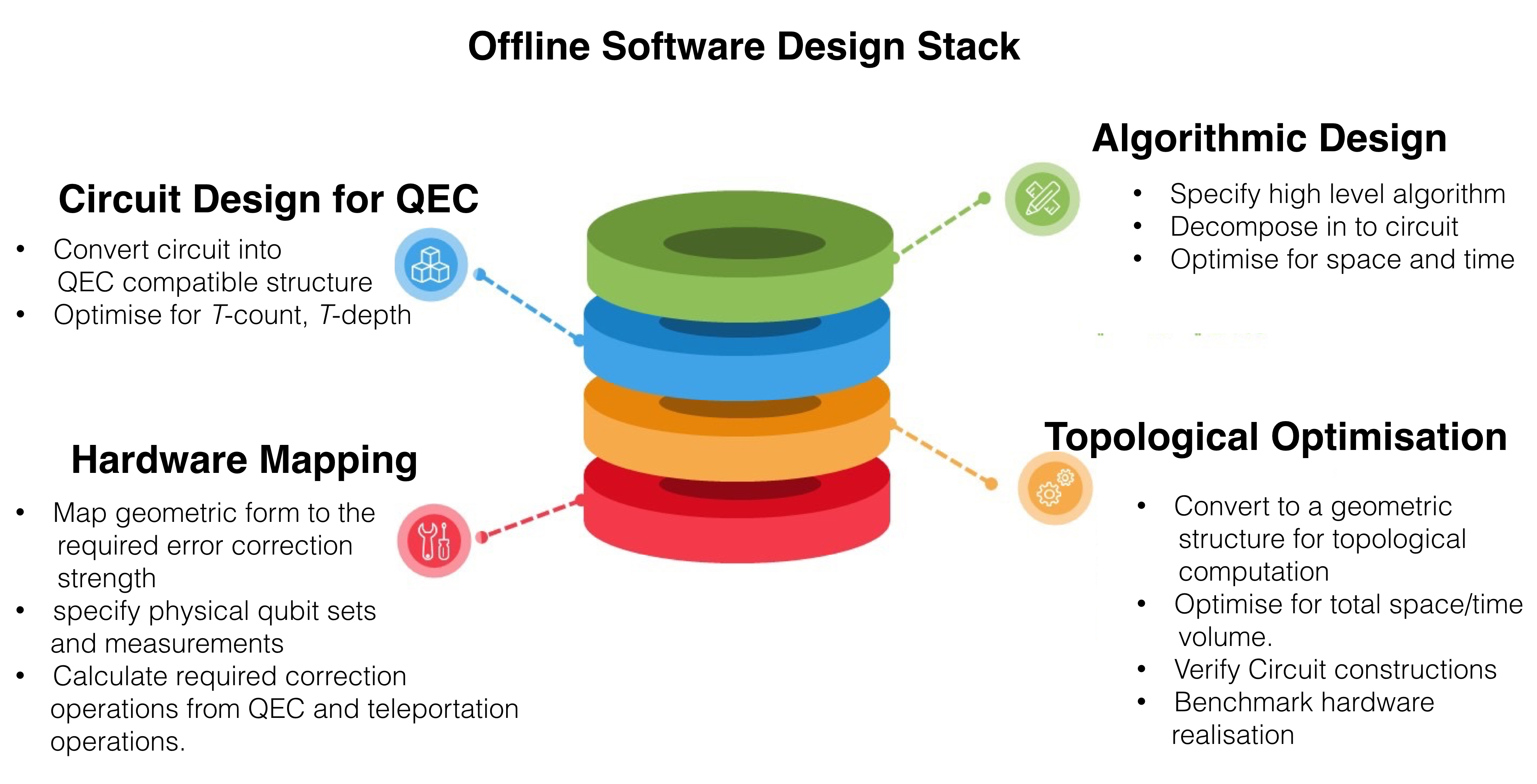}}
\end{center}
\caption{{\bf Offline design stack.} There are multiple stages to
  compiling and optimising a topological quantum circuit.  
  }
\label{fig:offline}
\end{figure*}

\begin{figure*}[ht!]
\begin{center}
\resizebox{\linewidth}{!}{\includegraphics{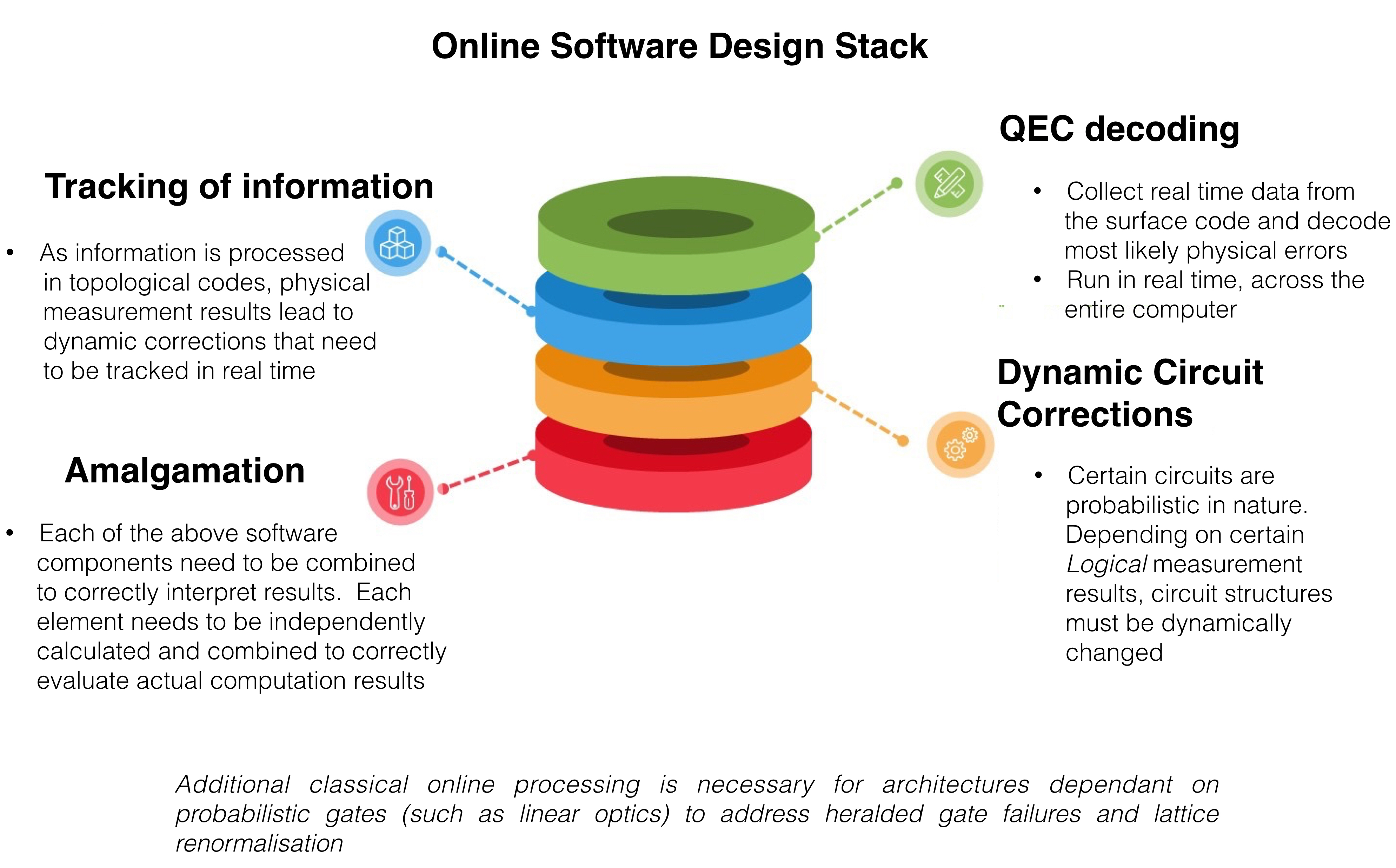}}
\end{center}
\caption{{\bf Online design stack.} As with offline computation, there
  are multiple layers to online classical processing for a large scale
  quantum computer.  Each of these components must {\em keep up} with
  the physical clock speed of a quantum computer.}
\label{fig:online}
\end{figure*}

Topological coding models of error corrected computation are software
based~\fewcites{devitt2014classical}\manycites{devitt2014classical}.
Enacting quantum algorithms is a function of switching on and off
sections of the computer in accordance with the overlying algorithm,
while error correction is a continuous process of extracting syndrome
information and decoding it to determine where physical errors have
occurred.  A large-scale quantum computer will require extensive
classical computational resources to operate.  These resources are
divided into two main categories: Offline control and Online control.
The various elements of both are illustrated in Figures
\ref{fig:offline} and \ref{fig:online}.

Offline control is the compilation and optimization of fault-tolerant
quantum circuits prior to turning on the computer
\manycites{G06,WS14,GLRSV13,WBCB14,PPND15,PDF16}.  These software elements
are needed to translate an abstract algorithm into gate sequences
compatible with fault-tolerant error correction and to translate these
gate lists into an appropriate control structure for the topological
codes.  At each stage of this offline compilation, circuits and
topological structures must be optimized for both physical qubits and
computational time \manycites{MDMN08,VM06,WS14} and optimized structures
must be verified against the desired computational specification
\manycites{SACCM06,PDNP14}.

Online control is the set of classical software packages that run in
tandem with the quantum computer.  They are primarily responsible for
dynamic error decoding \manycites{GP10,GP14,F15+} and the mapping of the
compiled circuit into the physical control and signals to the hardware
itself \manycites{PDNP14+}.  Online control software will require extremely
fast operation over a large dataset.  The algorithms must be able to
keep up with the physical clock rate of the quantum hardware (which
for 3rd generation machines will be in the GHz range) and will need to
operate on qubit arrays consisting of millions (perhaps billions) of
qubits \manycites{DSMN13}.  Consequently, the scaling properties of these
algorithms are a serious concern and need to be developed further.

We cannot operate a quantum computer without these packages and
appropriate benchmarking of quantum algorithms cannot be performed
without a fully developed compiler and software stack.  While there is
some work on compiling and benchmarking topological quantum circuits
\manycites{DSMN13}, and a large amount of work related to higher level
software languages and circuit compilers \manycites{WS14,GLRSV13,G14} there
is still much work to be done to optimize functional topological
circuits to the level known to be theoretically possible \manycites{F12+}
and to accurately determine qubit counts and computational time for
useful quantum algorithms.

\section{Networks and Distributed Applications}
\label{sec:net}

As already noted, the demand for scalable systems with high capacity
forces us into the realm of multicomputers, groups of smaller
computers connected via some form of system area network (SAN).
Specific hardware platforms have been proposed, building on ion traps,
quantum dots, or NV diamond, which offer good optical
connections~\manycites{lim05:_repeat_until_success,oi06:_dist-ion-trap-qec,jiang07:PhysRevA.76.062323,kim09:_integ_optic_ion_trap,van-meter10:dist_arch_ijqi,duan:RevModPhys.82.1209,nickerson2013topological,PhysRevA.89.022317}.
To make use of such systems, we must split an ordinarily monolithic
computation into pieces for distributed
computation~\manycites{van-meter06:thesis}.

Metropolitan area and wide area networks are also under
development~\manycites{van-meter14:_quantum_networking}.  To take advantage
of such networks, we need distributed quantum applications, which we
can divide into three categories: distributed numeric
computation~\manycites{broadbent:1412717,buhrman03:_dist_qc}, cryptographic
functions~\manycites{Wiesner83:quantum-money,bennett:bb84,ekert1991qcb,ben-or2005fast,cleve1999share},
and sensor or cybernetic
services~\manycites{jozsa2000qcs,chuang2000qclk,giovannetti2001quantum,hwang2002eqc,PhysRevLett.91.217905,RevModPhys.79.555,komar14:_clock_qnet,1367-2630-16-6-063040}.
Blind quantum computation allows client-server computation in which
the server cannot determine the input data, algorithm used, or output
data.  Cryptographic functions include secret key generation,
Byzantine agreement, and secret sharing.  Sensor uses include
high-precision interferometry and clock synchronization.

\section{Conclusion}

Over the last decade, experimental groups in a variety of
implementation technologies have met the DiVincenzo criteria,
including the error correction threshold at which applying error
correction removes more errors than it introduces.  In parallel,
theorists have analyzed multicomputer architectures and developed in
depth topological methods for error correction.  The process of
combining these concepts with experimental work is just beginning.

The quest for the smallest economically viable quantum computer is
therefore entering a new phase.  We reprise the question of Van Meter
and Horsman~\manycites{van-meter13:_blueprint}:

\begin{quotation}
  When will the first paper appear in {\em Science} or {\em Nature} in
  which the point is the results of a quantum computation, rather than
  the machine itself?  That is, when will a quantum computer {\em do}
  science, rather than {\em be} science?
\end{quotation}

While significant problems remain to be solved, the fundamental
questions about how to build a quantum computer now have positive
answers.  It is clear that quantum computing is now moving from the
research phase into the engineering phase, and we hope that such a
paper will appear within a decade.

{\bf Rodney Van Meter} holds degrees from the California Institute of
Technology, the University of Southern California, and Keio
University.  His research interests include storage systems,
networking, and post-Moore's Law computer architecture.  He has held
positions in both industry and academia in the U.S. and Japan.  He is
now an Associate Professor of Environment and Information Studies at
Keio University's Shonan Fujisawa Campus.

{\bf Simon J. Devitt} holds degrees from the University of Melbourne,
Australia.  His research is focused on large-scale architecture
designs for quantum computation and communications systems and
software compilation and optimisation for topological quantum
computing.  He is currently a senior research scientist at the center
for emergent matter sciences, Riken, Japan.

{\bf Acknowledgments.}  This work was supported by the Japan Society
for the Promotion of Science (JSPS) through KAKENHI Kiban B 16H02812
and the JSPS grant for challenging exploratory research.

\bibliographystyle{unsrt}
\bibliography{paper-reviews.bib,ieee.bib} 

\begin{thebibliography}{100}

\bibitem{nielsen-chuang:qci}
Michael~A. Nielsen and Isaac~L. Chuang.
\newblock {\em Quantum Computation and Quantum Information}.
\newblock Cambridge University Press, 2000.

\bibitem{van-meter14:_quantum_networking}
Rodney Van{ }Meter.
\newblock {\em Quantum Networking}.
\newblock Wiley-ISTE, April 2014.

\bibitem{deutsch85:_quant_church_turing}
David Deutsch.
\newblock Quantum theory, the {Church-Turing} principle, and the universal
  quantum computer.
\newblock {\em Proc. Royal Soc. London A}, 400:97--117, 1985.

\bibitem{feynman:_simul_physic_comput}
Richard~P. Feynman.
\newblock Simulating physics with computers.
\newblock In Anthony J.~G. Hey, editor, {\em Feynman and Computation}. Westview
  Press, 2002.

\bibitem{bennett:strengths}
C.~H. Bennett, E.~Bernstein, G.~Brassard, and U.~Vazirani.
\newblock Strengths and weaknesses of quantum computing.
\newblock {\em SIAM J. Comput.}, 26(5):1510--1523, 1997.
\newblock http://arXiv.org/quant-ph/9701001.

\bibitem{bernstein97:_quant_complexity}
Ethan Bernstein and Umesh Vazirani.
\newblock Quantum complexity theory.
\newblock {\em SIAM J. Computing}, 26(5):1411--1473, 1997.

\bibitem{mosca2008quantum}
M.~Mosca.
\newblock {Quantum algorithms}.
\newblock {\em Arxiv preprint arXiv:0808.0369}, 2008.

\bibitem{bacon10:_recent_progress}
Dave Bacon and Wim van{ }Dam.
\newblock Recent progress in quantum algorithms.
\newblock {\em Communications of the ACM}, 53(2):84--93, February 2010.

\bibitem{montanaro2015:qualgo-qi}
Ashley Montanaro.
\newblock Quantum algorithms: an overview.
\newblock {\em npj Quantum Information}, 2:15023, 2016.

\bibitem{divincenzo2000piq}
D.P. DiVincenzo.
\newblock {The physical implementation of quantum computation}.
\newblock {\em Fortschritte der Physik}, 48(9-11):771--783, 2000.

\bibitem{unruh95:_maintaining-coherence}
W.G. Unruh.
\newblock Maintaining coherence in quantum computers.
\newblock {\em Physical Review A}, 51:992--997, 1995.

\bibitem{K03}
A.~Kitaev.
\newblock {Fault-tolerant quantum computation by anyons}.
\newblock {\em Ann. Phys.}, 303:2, 2003.

\bibitem{FMMC12}
A.G. Fowler, M.~Mariantoni, J.M. Martinis, and A.N. Cleland.
\newblock {Surface codes: Towards practical large-scale quantum computation}.
\newblock {\em Phys. Rev. A.}, 86:032324, 2012.

\bibitem{RHG06}
R.~Raussendorf, J.~Harrington, and K.~Goyal.
\newblock {A Fault-tolerant one way quantum computer}.
\newblock {\em Ann. Phys.}, 321:2242, 2006.

\bibitem{DFSG08}
S.J. Devitt, A.G. Fowler, A.M. Stephens, A.D. Greentree, L.C.L. Hollenberg,
  W.J. Munro, and K.~Nemoto.
\newblock {Architectural design for a topological cluster state quantum
  computer}.
\newblock {\em New. J. Phys.}, 11:083032, 2009.

\bibitem{JMFMKLY10}
N.~Cody Jones, R.~Van Meter, A.G. Fowler, P.L. McMahon, J.~Kim, T.D. Ladd, and
  Y.~Yamamoto.
\newblock {A Layered Architecture for Quantum Computing Using Quantum Dots}.
\newblock {\em Phys. Rev. X.}, 2:031007, 2012.

\bibitem{YJG10}
N.Y. Yao, L.~Jiang, A.V. Gorshkov, P.C. Maurer, G.~Giedke, J.I. Cirac, and M.D.
  Lukin.
\newblock {Scalable Architecture for a Room Temperature Solid-State Quantum
  Information Processor}.
\newblock {\em Nature Communications}, 3:800, 2012.

\bibitem{NTDS13}
K.~Nemoto, M.~Trupke, S.J. Devitt, A.M. Stephens, K.~Buczak, T.~Nobauer, M.S.
  Everitt, J.~Schmiedmayer, and W.J. Munro.
\newblock {Photonic architecture for scalable quantum information processing in
  NV-diamond}.
\newblock {\em arXiv:1309.4277}, 2013.

\bibitem{MRRBMD14}
C.~Monroe, R.~Raussendorf, A.~Ruthven, K.R. Brown, P.~Maunz, L.-M. Duan, and
  J.~Kim.
\newblock {Large Scale Modular Quantum Computer Architecture with Atomic Memory
  and Photonic Interconnects}.
\newblock {\em Phys. Rev. A.}, 89:022317, 2014.

\bibitem{AMK15}
M.~Ahsan, R.~Van Meter, and J.~Kim.
\newblock {Designing a Million-Qubit Quantum Computer Using Resource
  Performance Simulator}.
\newblock {\em arxiv:1512.00796}, 2015.

\bibitem{LWFMDWH15}
B.~Lekitsch, S.~Weidt, A.G. Fowler, K.~M{\o o}lmer, S.J. Devitt, C.~Wunderlich,
  and W.K. Hensinger.
\newblock {Blueprint for a microwave ion trap quantum computer}.
\newblock {\em arxiv:1508.00420}, 2015.

\bibitem{LHMB15}
Y.~Li, P.C. Humphreys, G.J. Mendoza, and S.C. Benjamin.
\newblock {Resource Costs for Fault-Tolerant Linear Optical Quantum Computing}.
\newblock {\em Phys. Rev. X.}, 5:041007, 2015.

\bibitem{HPHHFRSH15}
C.D. Hill, E.~Peretz, S.J. Hile, M.G. House, M.~Fuechsle, S.~Rogge, M.Y.
  Simmons, and L.C.L. Hollenberg.
\newblock {A surface code quantum computer in silicon}.
\newblock {\em Science Advances}, 1(9):e1500707, 2015.

\bibitem{ONRMB16}
J.~O'Gorman, N.H. Nickerson, P.~Ross, J.J.L. Morton, and S.C. Benjamin.
\newblock {A silicon-based surface code quantum computer}.
\newblock {\em npj Quantum Information}, 2:16014, 2016.

\bibitem{S14}
A.M. Stephens.
\newblock {Fault-tolerant thresholds for quantum error correction with the
  surface code}.
\newblock {\em Phys. Rev. A.}, 89:022321, 2014.

\bibitem{CWCW13}
J.~Chen, L.~Wang, E.~Charbon, and B.~Wang.
\newblock {Programmable architecture for quantum computing}.
\newblock {\em Phys. Rev. A.}, 88(022311), 2013.

\bibitem{D14}
S.J. Devitt.
\newblock {Classical Control of Large-Scale Quantum Computers}.
\newblock {\em RC2014, Springer Lecture Notes on Computer Science (LNCS)},
  8507(26-39), 2014.

\bibitem{PPND15}
A.~Paler, I.~Polian, K.~Nemoto, and S.J. Devitt.
\newblock {A Compiler for Fault-Tolerant High Level Quantum Circuits}.
\newblock {\em arxiv:1509.02004}, 2015.

\bibitem{PDF16}
A.~Paler, S.J. Devitt, and A.G. Fowler.
\newblock {Synthesis of Arbitrary Quantum Circuits to Topological Assembly}.
\newblock {\em arxiv:1604.08621}, 2016.

\bibitem{van-meter10:dist_arch_ijqi}
Rodney Van{ }Meter, Thaddeus~D. Ladd, Austin~G. Fowler, and Yoshihisa Yamamoto.
\newblock Distributed quantum computation architecture using semiconductor
  nanophotonics.
\newblock {\em International Journal of Quantum Information}, 8:295--323, 2010.

\bibitem{van-meter13:_blueprint}
Rodney Van{ }Meter and Clare Horsman.
\newblock A blueprint for building a quantum computer.
\newblock {\em Communications of the ACM}, 53(10):84--93, October 2013.

\bibitem{LB13}
D.A. Lidar and T.A. Brun, editors.
\newblock {\em {Quantum Error Correction}}.
\newblock Cambridge University Press, 2013.

\bibitem{F15}
K.~Fujii.
\newblock {\em {Quantum Computation with Topological Codes. From Qubits to
  Topological Fault-Tolerance}}.
\newblock Springer -Verlag, Berlin ; New York, 2015.

\bibitem{horsman2012surface-njp}
C.~Horsman, A.G. Fowler, S.~Devitt, and R.~Van{ }Meter.
\newblock Surface code quantum computing by lattice surgery.
\newblock {\em New Journal of Physics}, 14:123011, 2012.

\bibitem{FD12}
A.G. Fowler and S.J. Devitt.
\newblock {A bridge to lower overhead quantum computation}.
\newblock {\em arXiv:1209.0510}, 2012.

\bibitem{ahsan2015designing}
Muhammad Ahsan, Rodney Van{ }Meter, and Jungsang Kim.
\newblock Designing a million-qubit quantum computer using a resource
  performance simulator.
\newblock {\em ACM Journal on Emerging Technologies in Computing Systems
  (JETC)}, 12(4):39, 2015.

\bibitem{ladd10:_quantum_computers}
T.D. Ladd, F.~Jelezko, R.~Laflamme, Y.~Nakamura, C.~Monroe, and J.L. O'Brien.
\newblock Quantum computers.
\newblock {\em Nature}, 464:45--53, March 2010.

\bibitem{van-meter06:thesis}
Rodney~Doyle Van{ }Meter{ }III.
\newblock {\em Architecture of a Quantum Multicomputer Optimized for {Shor's}
  Factoring Algorithm}.
\newblock PhD thesis, Keio University, 2006.
\newblock available as arXiv:quant-ph/0607065.

\bibitem{brown2016co}
KR~Brown, J~Kim, and C~Monroe.
\newblock Co-designing a scalable quantum computer with trapped atomic ions.
\newblock {\em arXiv preprint arXiv:1602.02840}, 2016.

\bibitem{CZ95}
J.I. Cirac and P.~Zoller.
\newblock {Quantum Computations with Cold Trapped Ions}.
\newblock {\em Phys. Rev. Lett.}, 74:4091, 1995.

\bibitem{MMKIW95}
C.~Monroe, D.M. Meekhof, B.E. King, W.M. Itano, and D.J. Wineland.
\newblock {Demonstration of a Fundamental quantum logic gate}.
\newblock {\em Phys. Rev. Lett.}, 75:4714, 1995.

\bibitem{LBMW03}
D.~Leibfried, R.~Blatt, C.~Monroe, and D.~Wineland.
\newblock {Quantum Dynamics of single trapped ions}.
\newblock {\em Rev. Mod. Phys.}, 75(281-324), 2003.

\bibitem{KMW02}
D.~Kielpinski, C.~Monroe, and D.J. Wineland.
\newblock {Architecture for a large-scale ion-trap quantum computer}.
\newblock {\em Nature (London)}, 417:709, 2002.

\bibitem{H06}
W.K. Hensinger, S.~Olmschenk, D.~Stick, D.~Hucul, M.~Yeo, M.~Acton,
  L.~Deslauriers, J.~Rabchuk, and C.~Munro.
\newblock { T-junction multi-zone ion trap array for two-dimensional ion
  shuttling, storage and manipulation}.
\newblock {\em Appl. Phys. Lett.}, 88:034101, 2006.

\bibitem{BOVABLW09}
R.B. Blakestad, C.~Ospelkaus, A.P. VanDevender, J.M. Amini, J.~Britton,
  D.~Leibfried, and D.J. Wineland.
\newblock {High-Fidelity Transport of Trapped-Ion Qubits through an X-Junction
  Trap Array}.
\newblock {\em Phys. Rev. Lett.}, 102:153002, 2009.

\bibitem{DMMMKM06}
L.-M. Duan, M.J. Madsen, D.L. Moehring, P.~Maunz, R.N.~Kohn Jr., and C.~Monroe.
\newblock {Probabilistic quantum gates between remote atoms through
  interference of optical frequency qubits}.
\newblock {\em Phys. Rev. A.}, 73:062324, 2006.

\bibitem{hucul2014modular}
D~Hucul, IV~Inlek, G~Vittorini, C~Crocker, S~Debnath, SM~Clark, and C~Monroe.
\newblock Modular entanglement of atomic qubits using photons and phonons.
\newblock {\em Nature Physics}, 11:37--42, 2014.

\bibitem{B14}
R.~Barends, J.~Kelly, A.~Megrant, A.~Veitia, D.~Sank, E.~Jeffrey, T.C. White,
  J.~Mutus, A.G. Fowler, B.~Campbell, Y.~Chen, Z.~Chen, B.~Chiaro,
  A.~Dunsworth, C.~Neill, P.~O`Malley, P.~Roushan, A.~Vainsencher, J.~Wenner,
  A.N. Korotkov, A.N. Cleland, and J.M. Martinis.
\newblock {Logic gates at the surface code threshold: Superconducting qubits
  poised for fault-tolerant quantum computing}.
\newblock {\em Nature (London)}, 508:500--503, 2014.

\bibitem{D09}
D.P. DiVincenzo.
\newblock {Fault-tolerant architectures for superconducting qubits}.
\newblock {\em Phys. Scr.}, T137, 2009.

\bibitem{B16}
T.~Brecht, W.~Pfaff, C.~Wang, Y.~Chu, L.~Frunzio, M.H. Devoret, and R.J.
  Schoelkopf.
\newblock {Multilayer microwave integrated quantum circuits for scalable
  quantum computing}.
\newblock {\em npj Quantum Information}, 2:16002, 2016.

\bibitem{C15}
A.~D. Corcoles, Easwar Magesan, Srikanth~J. Srinivasan, Andrew~W. Cross,
  M.~Steffen, Jay~M. Gambetta, and Jerry~M. Chow.
\newblock Demonstration of a quantum error detection code using a square
  lattice of four superconducting qubits.
\newblock {\em Nat Commun}, 6, 04 2015.

\bibitem{Bonneau2016}
Damien Bonneau, Joshua~W. Silverstone, and Mark~G. Thompson.
\newblock {\em Silicon Photonics III: Systems and Applications}, chapter
  Silicon Quantum Photonics, pages 41--82.
\newblock Springer Berlin Heidelberg, Berlin, Heidelberg, 2016.

\bibitem{KLM01}
E.~Knill, R.~Laflamme, and G.J. Milburn.
\newblock {A Scheme for Efficient Quantum Computation with linear optics}.
\newblock {\em Nature (London)}, 409:46, 2001.

\bibitem{OPWRB03}
J.L. O'Brien, G.J. Pryde, A.G. White, T.C. Ralph, and D.~Branning.
\newblock {Demonstration of an all optical quantum controlled-NOT gate}.
\newblock {\em Nature (London)}, 426:264, 2003.

\bibitem{POWBR04}
G.J. Pryde, J.L. O'Brien, A.G. White, S.D. Bartlett, and T.C. Ralph.
\newblock {Measuring a photonic qubit without destroying it}.
\newblock {\em Phys. Rev. Lett.}, 92:190402, 2004.

\bibitem{P08}
Alberto Politi, Martin~J. Cryan, John~G. Rarity, Siyuan Yu, and Jeremy~L.
  O'Brien.
\newblock Silica-on-silicon waveguide quantum circuits.
\newblock {\em Science}, 320(5876):646--649, 05 2008.

\bibitem{O07}
Jeremy~L. O'Brien.
\newblock Optical quantum computing.
\newblock {\em Science}, 318(5856):1567--1570, 12 2007.

\bibitem{O09}
Jeremy~L. O'Brien, Akira Furusawa, and Jelena Vuckovic.
\newblock Photonic quantum technologies.
\newblock {\em Nat Photon}, 3(12):687--695, 12 2009.

\bibitem{SB14}
SilverstoneJ. W., BonneauD., OhiraK., SuzukiN., YoshidaH., IizukaN., EzakiM.,
  NatarajanC. M., TannerM. G., HadfieldR. H., ZwillerV., MarshallG. D.,
  RarityJ. G., O'BrienJ. L., and ThompsonM. G.
\newblock On-chip quantum interference between silicon photon-pair sources.
\newblock {\em Nat Photon}, 8(2):104--108, 02 2014.

\bibitem{S15}
P.~Sibson, C.~Erven, M.~Godfrey, S.~Miki, T.~Yamashita, M.~Fujiwara, M.~Sasaki,
  H.~Terai, M.G. Tanner, C.M. Natarajan, R.H. Hadfield, J.L. O'Brien, and M.G.
  Thompson.
\newblock {Chip-based Quantum Key Distribution}.
\newblock {\em arxiv:1509.00768}, 2015.

\bibitem{SS15}
J.~W. Silverstone, R.~Santagati, D.~Bonneau, M.~J. Strain, M.~Sorel, J.~L.
  O/'Brien, and M.~G. Thompson.
\newblock Qubit entanglement between ring-resonator photon-pair sources on a
  silicon chip.
\newblock {\em Nat Commun}, 6, 08 2015.

\bibitem{KH13}
J.~E. Kennard, J.~P. Hadden, L.~Marseglia, I.~Aharonovich, S.~Castelletto,
  B.~R. Patton, A.~Politi, J.~C.~F. Matthews, A.~G. Sinclair, B.~C. Gibson,
  S.~Prawer, J.~G. Rarity, and J.~L. O'Brien.
\newblock On-chip manipulation of single photons from a diamond defect.
\newblock {\em Physical Review Letters}, 111(21):213603--, 11 2013.

\bibitem{SG16}
SomaschiN., GieszV., De~SantisL., LoredoJ. C., AlmeidaM. P., HorneckerG.,
  PortalupiS. L., GrangeT., Ant{\'o}nC., DemoryJ., G{\'o}mezC., SagnesI.,
  Lanzillotti-KimuraN. D., Lema{\'\i}treA., AuffevesA., WhiteA. G., LancoL.,
  and SenellartP.
\newblock Near-optimal single-photon sources in the solid state.
\newblock {\em Nat Photon}, 10(5):340--345, 05 2016.

\bibitem{CB10}
Julien Claudon, Joel Bleuse, Nitin~Singh Malik, Maela Bazin, Perine Jaffrennou,
  Niels Gregersen, Christophe Sauvan, Philippe Lalanne, and Jean-Michel Gerard.
\newblock A highly efficient single-photon source based on a quantum dot in a
  photonic nanowire.
\newblock {\em Nat Photon}, 4(3):174--177, 03 2010.

\bibitem{GM13}
O.~Gazzano, S.~Michaelis~de Vasconcellos, C.~Arnold, A.~Nowak, E.~Galopin,
  I.~Sagnes, L.~Lanco, A.~Lema{\^\i}tre, and P.~Senellart.
\newblock Bright solid-state sources of indistinguishable single photons.
\newblock {\em Nat Commun}, 4:1425, 02 2013.

\bibitem{MZ11}
Xiao-song Ma, Stefan Zotter, Johannes Kofler, Thomas Jennewein, and Anton
  Zeilinger.
\newblock Experimental generation of single photons via active multiplexing.
\newblock {\em Physical Review A}, 83(4):043814--, 04 2011.

\bibitem{MSM15}
G.J. Mendoza, R.~Santagati, J.~Munns, E.~Hemsley, M.~Piekarek, E.~Martin-Lopez,
  G.D. Marshall, D.~Bonneau, M.G. Thompson, and J.L. O'Brien.
\newblock {Active Temporal Multiplexing of Photons}.
\newblock {\em arXiv:1503.01215}, 2015.

\bibitem{BMO15}
Damien Bonneau, Gabriel~J Mendoza, Jeremy~L O'Brien, and Mark~G Thompson.
\newblock Effect of loss on multiplexed single-photon sources.
\newblock {\em New Journal of Physics}, 17(4):043057, 2015.

\bibitem{LBS10}
Ying Li, Sean~D. Barrett, Thomas~M. Stace, and Simon~C. Benjamin.
\newblock Fault tolerant quantum computation with nondeterministic gates.
\newblock {\em Physical Review Letters}, 105(25):250502--, 12 2010.

\bibitem{MSBR15}
Mercedes Gimeno-Segovia, Pete Shadbolt, Dan~E. Browne, and Terry Rudolph.
\newblock From three-photon greenberger-horne-zeilinger states to ballistic
  universal quantum computation.
\newblock {\em Physical Review Letters}, 115(2):020502--, 07 2015.

\bibitem{G11}
W.~P. Grice.
\newblock Arbitrarily complete bell-state measurement using only linear optical
  elements.
\newblock {\em Physical Review A}, 84(4):042331--, 10 2011.

\bibitem{greentree16:diamond}
Andrew~D Greentree.
\newblock Nanodiamonds in {Fabry-Perot} cavities: a route to scalable quantum
  computing.
\newblock {\em New Journal of Physics}, 18(2):021002, 2016.

\bibitem{PG08}
Steven Prawer and Andrew~D. Greentree.
\newblock Diamond for quantum computing.
\newblock {\em Science}, 320(5883):1601--1602, 06 2008.

\bibitem{LJLNMO10}
T.D. Ladd, F.~Jelezko, R.~Laflamme, Y.~Nakamura, C.~Monroe, and J.L. O'Brien.
\newblock {Quantum Computers}.
\newblock {\em Nature (London)}, 464:45--53, 2010.

\bibitem{TC10}
E.~Togan, Y.~Chu, A.~S. Trifonov, L.~Jiang, J.~Maze, L.~Childress, M.~V.~G.
  Dutt, A.~S. Sorensen, P.~R. Hemmer, A.~S. Zibrov, and M.~D. Lukin.
\newblock Quantum entanglement between an optical photon and a solid-state spin
  qubit.
\newblock {\em Nature}, 466(7307):730--734, 08 2010.

\bibitem{NB10}
Philipp Neumann, Johannes Beck, Matthias Steiner, Florian Rempp, Helmut Fedder,
  Philip~R. Hemmer, J{\"o}rg Wrachtrup, and Fedor Jelezko.
\newblock Single-shot readout of a single nuclear spin.
\newblock {\em Science}, 329(5991):542--544, 07 2010.

\bibitem{RC11}
Lucio Robledo, Lilian Childress, Hannes Bernien, Bas Hensen, Paul F.~A.
  Alkemade, and Ronald Hanson.
\newblock High-fidelity projective read-out of a solid-state spin quantum
  register.
\newblock {\em Nature}, 477(7366):574--578, 09 2011.

\bibitem{MK12}
P.~C. Maurer, G.~Kucsko, C.~Latta, L.~Jiang, N.~Y. Yao, S.~D. Bennett,
  F.~Pastawski, D.~Hunger, N.~Chisholm, M.~Markham, D.~J. Twitchen, J.~I.
  Cirac, and M.~D. Lukin.
\newblock Room-temperature quantum bit memory exceeding one second.
\newblock {\em Science}, 336(6086):1283--1286, 06 2012.

\bibitem{BH13}
H.~Bernien, B.~Hensen, W.~Pfaff, G.~Koolstra, M.~S. Blok, L.~Robledo, T.~H.
  Taminiau, M.~Markham, D.~J. Twitchen, L.~Childress, and R.~Hanson.
\newblock Heralded entanglement between solid-state qubits separated by three
  metres.
\newblock {\em Nature}, 497(7447):86--90, 05 2013.

\bibitem{DJ13}
F.~Dolde, I.~Jakobi, B.~Naydenov, N.~Zhao, S.~Pezzagna, C.~Trautmann,
  J.~Meijer, P.~Neumann, F.~Jelezko, and J.~Wrachtrup.
\newblock Room-temperature entanglement between single defect spins in diamond.
\newblock {\em Nat Phys}, 9(3):139--143, 03 2013.

\bibitem{HBD15}
B.~Hensen, H.~Bernien, A.~E. Dreau, A.~Reiserer, N.~Kalb, M.~S. Blok,
  J.~Ruitenberg, R.~F.~L. Vermeulen, R.~N. Schouten, C.~Abellan, W.~Amaya,
  V.~Pruneri, M.~W. Mitchell, M.~Markham, D.~J. Twitchen, D.~Elkouss,
  S.~Wehner, T.~H. Taminiau, and R.~Hanson.
\newblock Loophole-free bell inequality violation using electron spins
  separated by 1.3 kilometres.
\newblock {\em Nature}, 526(7575):682--686, 10 2015.

\bibitem{ZS11}
X.~Zhu, S.~Saito, A.~Kemp, K.~Kakuyanagi, S.~Karimoto, H.~Nakano, W.~J. Munro,
  Y.~Tokura, M.S. Everitt, K.~Nemoto, M.~Kasu, N.~Mizuochi, and K.~Semba.
\newblock {Coherent coupling of a superconducting flux qubit to an electron
  spin ensemble in diamond}.
\newblock {\em Nature (London)}, 478:221--224, 2011.

\bibitem{zwanenburg:RevModPhys.85.961}
Floris~A. Zwanenburg, Andrew~S. Dzurak, Andrea Morello, Michelle~Y. Simmons,
  Lloyd C.~L. Hollenberg, Gerhard Klimeck, Sven Rogge, Susan~N. Coppersmith,
  and Mark~A. Eriksson.
\newblock Silicon quantum electronics.
\newblock {\em Rev. Mod. Phys.}, 85:961--1019, Jul 2013.

\bibitem{reed:PhysRevLett.116.110402}
M.~D. Reed, B.~M. Maune, R.~W. Andrews, M.~G. Borselli, K.~Eng, M.~P. Jura,
  A.~A. Kiselev, T.~D. Ladd, S.~T. Merkel, I.~Milosavljevic, E.~J. Pritchett,
  M.~T. Rakher, R.~S. Ross, A.~E. Schmitz, A.~Smith, J.~A. Wright, M.~F. Gyure,
  and A.~T. Hunter.
\newblock Reduced sensitivity to charge noise in semiconductor spin qubits via
  symmetric operation.
\newblock {\em Phys. Rev. Lett.}, 116:110402, Mar 2016.

\bibitem{LD98}
D.~Loss and D.P. DiVincenzo.
\newblock {Quantum Computation with quantum dots}.
\newblock {\em Phys. Rev. A.}, 57:120, 1998.

\bibitem{VD14}
M.~Veldhorst, J.C.C. Hwang, C.H. Yang, A.W. Leenstra, B.~de~Ronde, J.P.
  Dehollain, J.T. Muhonen, F.E. Hudson, K.M. Itoh, A.~Morello, and A.S. Dzurak.
\newblock {An addressable quantum dot qubit with fault-tolerant control
  fidelity}.
\newblock {\em Nature Nanotechnology}, 9:981, 2014.

\bibitem{TT16}
K.~Takeda, J.~Kamioka, T.~Otsuka, J.~Yoneda, T.~Nakajima, M.R. Delbecq,
  S.~Amaha, G.~Allison, T.~Kodera, S.~Oda, and S.~Tarucha.
\newblock {A fault-tolerant addressable spin qubit in a natural silicon quantum
  dot}.
\newblock {\em arXiv:1602.07833}, 2016.

\bibitem{VD15}
M.~Veldhorst, C.H. Yang, J.C.C. Hwang, W.~Huang, J.P. Dehollain, J.T. Muhonen,
  S.~Simmons, A.~Laucht, F.E. Hudson, K.M. Itoh, A.~Morello, and A.S. Dzurak.
\newblock {A Two Qubit Logic Gate in Silicon}.
\newblock {\em Nature (London)}, 526:410--414, 2015.

\bibitem{Hill:e1500707}
Charles~D. Hill, Eldad Peretz, Samuel~J. Hile, Matthew~G. House, Martin
  Fuechsle, Sven Rogge, Michelle~Y. Simmons, and Lloyd C.~L. Hollenberg.
\newblock A surface code quantum computer in silicon.
\newblock {\em Science Advances}, 1(9), 2015.

\bibitem{ZDM13}
Floris~A. Zwanenburg, Andrew~S. Dzurak, Andrea Morello, Michelle~Y. Simmons,
  Lloyd C.~L. Hollenberg, Gerhard Klimeck, Sven Rogge, Susan~N. Coppersmith,
  and Mark~A. Eriksson.
\newblock Silicon quantum electronics.
\newblock {\em Reviews of Modern Physics}, 85(3):961--1019, 07 2013.

\bibitem{K98}
B.E. Kane.
\newblock {A Silicon-Based nuclear spin Quantum Computer}.
\newblock {\em Nature (London)}, 393:133, 1998.

\bibitem{HGFW06}
L.C.L. Hollenberg, A.D. Greentree, A.G. Fowler, and C.J. Wellard.
\newblock {Two-Dimensional Architectures for Donor-Based Quantum Computing}.
\newblock {\em Phys. Rev. B.}, 74:045311, 2006.

\bibitem{DHJ03}
A.~S. Dzurak, L.~C.~L. Hollenberg, D.~N. Jamieson, F.~E. Stanley, C.~Yang,
  T.~M. Buhler, V.~Chan, D.~J. Reilly, C.~Wellard, A.~R. Hamilton, C.~I. Pakes,
  A.~G. Ferguson, E.~Gauja, S.~Prawer, G.~J. Milburn, and R.~G. Clark.
\newblock {Charge-based silicon quantum computer architectures using controlled
  single-ion implantation}.
\newblock {\em arXiv:cond-mat/0306265}, 2003.

\bibitem{SCS03}
S.~R. Schofield, N.~J. Curson, M.~Y. Simmons, F.~J. Rue{\ss}, T.~Hallam,
  L.~Oberbeck, and R.~G. Clark.
\newblock Atomically precise placement of single dopants in si.
\newblock {\em Physical Review Letters}, 91(13):136104--, 09 2003.

\bibitem{BMR13}
H.~B{\"u}ch, S.~Mahapatra, R.~Rahman, A.~Morello, and M.~Y. Simmons.
\newblock Spin readout and addressability of phosphorus-donor clusters in
  silicon.
\newblock {\em Nat Commun}, 4, 06 2013.

\bibitem{MDJ14}
Juha~T. Muhonen, Juan~P. Dehollain, Arne Laucht, Fay~E. Hudson, Rachpon Kalra,
  Takeharu Sekiguchi, Kohei~M. Itoh, David~N. Jamieson, Jeffrey~C. McCallum,
  Andrew~S. Dzurak, and Andrea Morello.
\newblock Storing quantum information for 30 seconds in a nanoelectronic
  device.
\newblock {\em Nat Nano}, 9(12):986--991, 12 2014.

\bibitem{PTK13}
Jarryd~J. Pla, Kuan~Y. Tan, Juan~P. Dehollain, Wee~H. Lim, John J.~L. Morton,
  Floris~A. Zwanenburg, David~N. Jamieson, Andrew~S. Dzurak, and Andrea
  Morello.
\newblock High-fidelity readout and control of a nuclear spin qubit in silicon.
\newblock {\em Nature}, 496(7445):334--338, 04 2013.

\bibitem{PTDJ12}
Jarryd~J. Pla, Kuan~Y. Tan, Juan~P. Dehollain, Wee~H. Lim, John J.~L. Morton,
  David~N. Jamieson, Andrew~S. Dzurak, and Andrea Morello.
\newblock A single-atom electron spin qubit in silicon.
\newblock {\em Nature}, 489(7417):541--545, 09 2012.

\bibitem{SSM16}
Juan~P. Dehollain, Stephanie Simmons, Juha~T. Muhonen, Rachpon Kalra, Arne
  Laucht, Fay Hudson, Kohei~M. Itoh, David~N. Jamieson, Jeffrey~C. McCallum,
  Andrew~S. Dzurak, and Andrea Morello.
\newblock Bell's inequality violation with spins in silicon.
\newblock {\em Nat Nano}, 11(3):242--246, 03 2016.

\bibitem{das-sarma2015:majorana-npjqi}
Sankar~Das Sarma, Michael Freedman, and Chetan Nayak.
\newblock Majorana zero modes and topological quantum computation.
\newblock {\em npj Quantum Information}, 1:15001, 2015.

\bibitem{FKLW03}
M.H. Freedman, A.~Kitaev, M.J. Larsen, and Z.~Wang.
\newblock {Topological quantum computation}.
\newblock {\em Bull. Amer. Math. Soc.}, 40:31--38, 2003.

\bibitem{NSS08}
Chetan Nayak, Steven~H. Simon, Ady Stern, Michael Freedman, and Sankar
  Das~Sarma.
\newblock Non-abelian anyons and topological quantum computation.
\newblock {\em Reviews of Modern Physics}, 80(3):1083--1159, 09 2008.

\bibitem{SDN15}
Sankar~Das Sarma, Michael Freedman, and Chetan Nayak.
\newblock Majorana zero modes and topological quantum computation.
\newblock {\em Npj Quantum Information}, 1:15001 EP --, 10 2015.

\bibitem{MZFK12}
V.~Mourik, K.~Zuo, S.M. Frolov, S.R. Plissard, E.P.A.M. Bakkers, and L.P.
  Kouwenhoven.
\newblock {Signatures of Majorana Fermions in Hybrid
  Superconductor-Semiconductor Nanowire Devices}.
\newblock {\em Science}, 336:1003--1007, 2012.

\bibitem{AHM16}
S.M. Albrecht, A.P. Higginbotham, M.~Madsen, F.~Kuemmeth, T.S. Jespersen,
  J.~Nygard, P.~Krogstrup, and C.M. Marcus.
\newblock {Exponential protection of zero modes in Majorana islands}.
\newblock {\em Nature (London)}, 531:206--209, 2016.

\bibitem{devitt2014classical}
Simon~J Devitt.
\newblock Classical control of large-scale quantum computers.
\newblock In {\em Reversible Computation}, pages 26--39. Springer International
  Publishing, 2014.

\bibitem{G06}
S.J. Gay.
\newblock {Quantum Programming Languages}.
\newblock {\em Mathematical Structures in Computer Science}, 16(04):581--600,
  2006.

\bibitem{WS14}
D.~Wecker and K.M. Svore.
\newblock {LIQUi|>: A Software Design Architecture and Domain-Specific Language
  for Quantum Computing}.
\newblock {\em arXiv:1402.4467}, 2014.

\bibitem{GLRSV13}
A.S. Green, P.L. Lumsdaine, N.J. Ross, P.~Selinger, and B.~Valiron.
\newblock {Quipper: A Scalable Quantum Programming Language}.
\newblock {\em ACM SIGPLAN Notices}, 48(6):333--342, 2013.

\bibitem{WBCB14}
Dave Wecker, Bela Bauer, Bryan~K. Clark, Matthew~B. Hastings, and Matthias
  Troyer.
\newblock Gate-count estimates for performing quantum chemistry on small
  quantum computers.
\newblock {\em Physical Review A}, 90(2):022305--, 08 2014.

\bibitem{MDMN08}
D.~Maslov, G.~W. Dueck, D.~M. Miller, and C.~Negrevergne.
\newblock {Quantum Circuit Simplification and Level Compaction}.
\newblock {\em IEEE Transactions on Computer-Aided Design of Integrated
  Circuits and Systems}, 27(3):436--444, 2008.

\bibitem{VM06}
R.~Van Meter.
\newblock {Architecture of a Quantum Multicomputer Optimized for Shor's
  Factoring Algorithm}.
\newblock {\em quant-ph/0607065}, 2006.

\bibitem{SACCM06}
K.M. Svore, A.V. Aho, A.W. Cross, I.L. Chuang, and I.L. Markov.
\newblock {A Layered Software Architecture for Quantum Computing Design Tools}.
\newblock {\em Computer}, 39(1):74--83, 2006.

\bibitem{PDNP14}
A.~Paler, S.J. Devitt, K.~Nemoto, and I.~Polian.
\newblock {Cross-level Validation of Topological Quantum Circuits}.
\newblock {\em Springer Lecture Notes on Computer Science (LNCS)},
  8507:189--200, 2014.

\bibitem{GP10}
G.~Duclos-Cianci and D.~Poulin.
\newblock {Fast Decoders for Topological Quantum Codes}.
\newblock {\em Phys. Rev. Lett.}, 104:050504, 2010.

\bibitem{GP14}
G.~Duclos-Cianci and D.~Poulin.
\newblock {Fault-Tolerant Renormalization Group Decoded for Abelian Topological
  Codes}.
\newblock {\em Quant. Inf. Comp.}, 14:0721, 2014.

\bibitem{F15+}
A.G. Fowler.
\newblock {Minimum weight perfect matching of fault-tolerant topological
  quantum error correction in average $O(1)$ parallel time}.
\newblock {\em Quant. Inf. Comp.}, 15(0145-0158), 2015.

\bibitem{PDNP14+}
Alexandru Paler, Simon~J Devitt, Kae Nemoto, and Ilia Polian.
\newblock Mapping of topological quantum circuits to physical hardware.
\newblock {\em Scientific reports}, 4, 2014.

\bibitem{DSMN13}
S.J. Devitt, A.M. Stephens, W.J. Munro, and K.~Nemoto.
\newblock Requirements for fault-tolerant factoring on an atom-optics quantum
  computer.
\newblock {\em Nature Communications}, 4:2524, 2013.

\bibitem{G14}
V.~Gheorghiu.
\newblock {Quantum++ - A C++11 quantum computing library}.
\newblock {\em arXiv:1412.4704}, 2014.

\bibitem{F12+}
A.G. Fowler.
\newblock {Time Optimal quantum computation}.
\newblock {\em arxiv:1210.4626}, 2012.

\bibitem{lim05:_repeat_until_success}
Yuan~Liang Lim, Sean~D. Barrett, Almut Beige, Pieter Kok, and Leong~Chuan Kwek.
\newblock {Repeat-Until-Success} quantum computing using stationary and flying
  qubits.
\newblock {\em Physical Review Letters}, 95(3):30505, 2005.

\bibitem{oi06:_dist-ion-trap-qec}
Daniel K.~L. Oi, Simon~J. Devitt, and Lloyd C.~L. Hollenberg.
\newblock Scalable error correction in distributed ion trap computers.
\newblock {\em Physical Review A}, 74:052313, 2006.

\bibitem{jiang07:PhysRevA.76.062323}
Liang Jiang, Jacob~M. Taylor, Anders~S. S\o{}rensen, and Mikhail~D. Lukin.
\newblock Distributed quantum computation based on small quantum registers.
\newblock {\em Phys. Rev. A}, 76:062323, Dec 2007.

\bibitem{kim09:_integ_optic_ion_trap}
Jungsang Kim and Changsoon Kim.
\newblock Integrated optical approach to trapped ion quantum computation.
\newblock {\em Quantum Information and Computation}, 9(2), 2009.

\bibitem{duan:RevModPhys.82.1209}
L.-M. Duan and C.~Monroe.
\newblock \textit{Colloquium} : Quantum networks with trapped ions.
\newblock {\em Rev. Mod. Phys.}, 82:1209--1224, Apr 2010.

\bibitem{nickerson2013topological}
Naomi~H Nickerson, Ying Li, and Simon~C Benjamin.
\newblock Topological quantum computing with a very noisy network and local
  error rates approaching one percent.
\newblock {\em Nature communications}, 4:1756, 2013.

\bibitem{PhysRevA.89.022317}
C.~Monroe, R.~Raussendorf, A.~Ruthven, K.~R. Brown, P.~Maunz, L.-M. Duan, and
  J.~Kim.
\newblock Large-scale modular quantum-computer architecture with atomic memory
  and photonic interconnects.
\newblock {\em Phys. Rev. A}, 89:022317, Feb 2014.

\bibitem{broadbent:1412717}
Anne Broadbent and Alain Tapp.
\newblock Can quantum mechanics help distributed computing?
\newblock {\em SIGACT News}, 39(3):67--76, 2008.

\bibitem{buhrman03:_dist_qc}
Harry Buhrman and Hein R\"ohrig.
\newblock {\em Mathematical Foundations of Computer Science 2003}, chapter
  Distributed Quantum Computing, pages 1--20.
\newblock Springer-Verlag, 2003.

\bibitem{Wiesner83:quantum-money}
Stephen Wiesner.
\newblock Conjugate coding.
\newblock {\em SIGACT News}, 15(1):78--88, January 1983.

\bibitem{bennett:bb84}
C.~H. Bennett and G.~Brassard.
\newblock Quantum cryptography: Public key distribution and coin tossing.
\newblock In {\em Proc. IEEE International Conference on Computers, Systems,
  and Signal Processing}, pages 175--179. IEEE, December 1984.

\bibitem{ekert1991qcb}
A.K. Ekert.
\newblock Quantum cryptography based on {Bell's} theorem.
\newblock {\em Physical Review Letters}, 67(6):661--663, 1991.

\bibitem{ben-or2005fast}
M.~Ben-Or and A.~Hassidim.
\newblock {Fast quantum Byzantine agreement}.
\newblock In {\em Proceedings of the thirty-seventh annual ACM symposium on
  Theory of computing}, pages 481--485. ACM, 2005.

\bibitem{cleve1999share}
Richard Cleve, Daniel Gottesman, and Hoi-Kwong Lo.
\newblock How to share a quantum secret.
\newblock {\em Physical Review Letters}, 83(3):648--651, 1999.

\bibitem{jozsa2000qcs}
R.~Jozsa, D.S. Abrams, J.P. Dowling, and C.P. Williams.
\newblock Quantum clock synchronization based on shared prior entanglement.
\newblock {\em Physical Review Letters}, 85(9):2010--2013, 2000.

\bibitem{chuang2000qclk}
I.L. Chuang.
\newblock Quantum algorithm for distributed clock synchronization.
\newblock {\em Physical Review Letters}, 85(9):2006--2009, 2000.

\bibitem{giovannetti2001quantum}
Vittorio Giovannetti, Seth Lloyd, and Lorenzo Maccone.
\newblock Quantum-enhanced positioning and clock synchronization.
\newblock {\em Nature}, 412(6845):417--419, 2001.

\bibitem{hwang2002eqc}
WY~Hwang, D.~Ahn, SW~Hwang, and YD~Han.
\newblock Entangled quantum clocks for measuring proper-time difference.
\newblock {\em The European Physical Journal D-Atomic, Molecular and Optical
  Physics}, 19(1):129--132, 2002.

\bibitem{PhysRevLett.91.217905}
Terry Rudolph and Lov Grover.
\newblock Quantum communication complexity of establishing a shared reference
  frame.
\newblock {\em Phys. Rev. Lett.}, 91:217905, Nov 2003.

\bibitem{RevModPhys.79.555}
Stephen~D. Bartlett, Terry Rudolph, and Robert~W. Spekkens.
\newblock Reference frames, superselection rules, and quantum information.
\newblock {\em Rev. Mod. Phys.}, 79:555--609, Apr 2007.

\bibitem{komar14:_clock_qnet}
P.~K{\'o}m{\'a}r, E.M. Kessler, M.~Bishof, L.~Jiang, A.~S. S{\/o}rensen, and
  M.~D. Lukin.
\newblock A quantum network of clocks.
\newblock {\em Nature Physics}, June 2014.

\bibitem{1367-2630-16-6-063040}
Tanvirul Islam, Lo\"ick Magnin, Brandon Sorg, and Stephanie Wehner.
\newblock Spatial reference frame agreement in quantum networks.
\newblock {\em New Journal of Physics}, 16(6):063040, 2014.

\end{thebibliography}

\end{document}